\documentclass[aps,pra,reprint,superscriptaddress,footinbib,floatfix,showpacs]{revtex4-1}

\usepackage{graphicx}
\usepackage{amsmath}
\usepackage{amssymb}
\usepackage{amsfonts}
\usepackage{bm}

\graphicspath{{.}{./EPS/}}

\newcommand{\ket}[1]{\left | \, #1 \right \rangle}
\newcommand{\bra}[1]{\left \langle #1 \, \right |}
\newcommand{\av}[1]{\langle #1\rangle}

\newcommand* {\ee}{\ensuremath{\mathrm{e}}}

\newcommand*{\vek}[1]{{\ensuremath{\bm{\mathrm{#1}}}}}

\begin{document}

\title{Majorana fermions from Landau quantization in a
  superconductor--topological-insulator hybrid structure}

\author{Rakesh P. Tiwari}
\affiliation{Department of Physics, University of Basel, 
Klingelbergstrasse 82, CH-4056 Basel, Switzerland}
 
\author{U. Z\"ulicke}
\affiliation{School of Chemical and Physical Sciences and MacDiarmid
  Institute for Advanced Materials and Nanotechnology, Victoria
  University of Wellington, PO Box 600, Wellington 6140, New Zealand}

\author{C. Bruder}
\affiliation{Department of Physics, University of Basel, 
Klingelbergstrasse 82, CH-4056 Basel, Switzerland}

\date{\today}

\begin{abstract}

We show that the interplay of cyclotron motion and Andreev reflection
experienced by massless-Dirac-like charge carriers in
topological-insulator surface states generates a Majorana-particle
excitation. Based on an envelope-function description of the
Dirac-Andreev edge states, we discuss the kinematic properties of the
Majorana mode and find them to be possible to be tuned by changing the
superconductor's chemical potential and/or the magnitude of the
perpendicular magnetic field.  Our proposal opens up new possibilities
for studying Majorana fermions in a controllable setup.
\end{abstract}


\maketitle

\textit{Introduction\/}. -- In recent years, the possibility of
observing and exploiting Majorana quasiparticles in condensed matter
systems has attracted a great deal of
attention~\cite{wil09,ali12,bee13,lei12}. Originally these elusive
fermionic particles were proposed as real solutions of
the Dirac equation~\cite{maj37}. Potential condensed-matter systems
hosting realizations of these atypical fermions include the $\nu=5/2$
quantum-Hall state~\cite{moo91,ste10}, $p$-wave
superconductors~\cite{iva01} such as strontium ruthenate~\cite{mac03},
semiconductor-superconductor
heterostructures~\cite{sau10,sau10a,ali10} and the surface of
topological insulators~\cite{has10,qi11} among others.
An experimental signature of localized Majorana excitations are
zero-bias conductance anomalies~\cite{sen01,bol07,law09,fle10} that
may have been measured recently~\cite{mou12,das12}. 

A particularly interesting route towards realizing chiral Majorana
modes involves hybrid structures of a topological insulator (TI) and
an s-wave superconductor~\cite{fu08,fu09}.  The requirement of broken
time-reversal symmetry and gapped excitation spectrum for the surface
states in the TI could be fulfilled by proximity to a ferromagnetic
insulator~\cite{fu09} or Zeeman splitting due to a magnetic
field~\cite{fu08}. Here we pursue an alternative scenario where Landau
quantization of the surface states' orbital motion in a uniform
perpendicular magnetic field is the origin of a gap and breaking of
time-reversal symmetry.  Besides avoiding materials-science challenges
associated with fabrication of hybrid structures involving three
different kinds of materials, our setup also offers several new
features enabling the manipulation of the Majorana excitation's
properties.

\begin{figure}[b]
\includegraphics[scale=0.35]{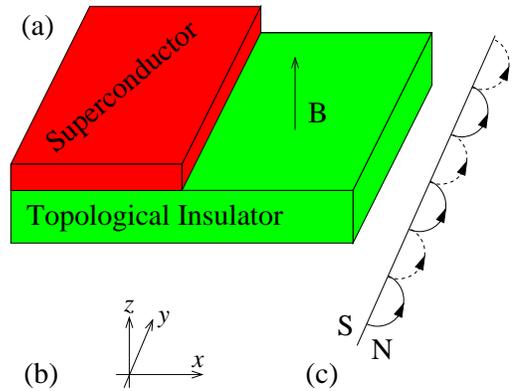}
\caption{\label{fig:setup} (Color online) (a) Schematics of the
  proposed sample layout. (b) The coordinate system.  (c) The
  semiclassical picture of Dirac-Andreev edge states (solid and dashed
  lines represent electrons and holes respectively).}

\end{figure}

Figure~\ref{fig:setup} illustrates the proposed sample geometry. The
massless-Dirac-like surface states of a bulk three-dimensional TI
material (e.g., Bi$_2$Se$_3$ \cite{has10}) occupy the
$xy$ plane. A planar contact with a superconductor in the half-plane
$x<0$ induces a pair potential for this part of the TI surface
(henceforth called the S region), while a uniform perpendicular
magnetic field $\vek{B} = B\hat{\vek{z}}$ is present in the half-plane
$x>0$ (the N region of the TI surface). We assume the magnetic field
to be fully screened from the S part and neglect Zeeman splitting
throughout~\footnote{A more detailed theoretical treatment of these
  effects can be developed along the lines of previous
  studies~\cite{zue01,gia05}.}. The interplay of cyclotron motion of
charge carriers in the N region with Andreev reflection from the
interface with the S region results in the formation of chiral
Dirac-Andreev edge states similar to the ones discussed previously for
an S--graphene hybrid structure~\cite{akh07}.  The quantum description
of these edge channels reveals one of them to be associated with a
chiral Majorana fermion mode with tunable velocity and
guiding-center-dependent electric charge.

\textit{Basic model for the S--TI hybrid structure\/}. -- To describe
single-particle excitations in the S--N heterostructure made from TI
surface states, we employ the Dirac-Bogoliubov-de~Gennes (DBdG)
equation~\cite{akh07,fu08}
\begin{equation}
\left( \begin{array}{cc}
H_{\text{D}}(\vek{r})-\mu & \Delta(\vek{r})\,\sigma_0 \\ 
\Delta^{\ast}(\vek{r})\, \sigma_0 &
\mu-T H_{\text{D}}(\vek{r})\, T^{-1} \end{array}\right) 
\Psi(\vek{r}) = \varepsilon \Psi(\vek{r}) \: ,
\label{eq:hamiltonian} 
\end{equation}
with the pair potential $\Delta(\vek{r})=\ee^{i\phi}
\Delta_0\,\Theta(-x)$ finite only in the S region,
$\sigma_0\equiv\openone_{2\times 2}$ denoting the two-dimensional
identity matrix, and $H_{\text{D}}(\vek{r}) = v_{\text{F}} [\vek{p} +
  e\vek{A}(\vek{r})] \cdot \vek{\sigma}$ being the massless-Dirac
Hamiltonian for the TI surface states. The position $\vek{r}\equiv (x,
y)$ and momentum $\vek{p}\equiv -i\hbar (\partial_x, \partial_y)$ are
restricted to the TI surface, $\vek{\sigma}$ is the vector of
Pauli matrices acting in spin space. Furthermore, $T$ denotes the
time-reversal operator, $-e$ the electron charge, and ${\bf A}$ the
vector potential associated with the magnetic field $\vek{B}=
\vek{\nabla}\times\vek{A}$. The excitation energy $\varepsilon$ is
measured relative to the chemical potential $\mu$ of the
superconductor, with the absolute zero of energy set to be at the
Dirac (i.e., neutrality) point of the TI surface states. The wave
function $\Psi$ from Eq.~(\ref{eq:hamiltonian}) is a spinor in
Dirac-Nambu space, which can be expressed explicitly in terms of
spin-resolved amplitudes as $\Psi=(u_{\uparrow}, u_{\downarrow},
v_{\downarrow},-v_{\uparrow})^{\text{T}}$.

\begin{figure}[b]
\includegraphics[angle=0,scale=1.2]{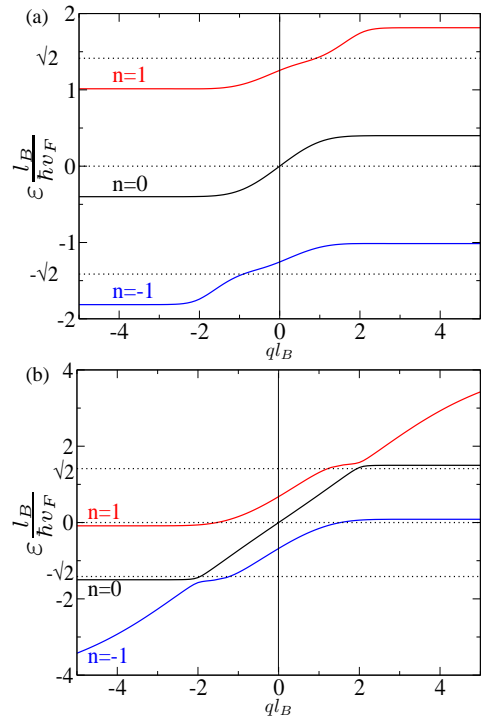}
\caption{\label{fig:envsq} (Color online) Dispersion relation
  $\varepsilon_{nq}$ of single-particle excitations at the S--N
  junction on a TI's surface and subject to a strong perpendicular
  magnetic field. Landau levels with $n=0, \pm1$ are shown for
  $\Delta_0 = 5 \hbar v_{\text{F}}/l_B$ and $\mu=0.4\hbar
  v_{\text{F}}/l_B$ [panel (a)], $\mu=1.5 \hbar v_{\text{F}}/l_B$
  [panel (b)], and $\mu =0$ (dotted
  lines). $l_B\equiv\sqrt{\hbar/|eB|}$ is the magnetic length.}
\end{figure}

To describe the uniform perpendicular magnetic field in the N region,
we adopt the Landau gauge $\vek{A}=B\, x\, \hat{\vek{y}}$. Then the
momentum $\hbar q$ parallel to the interface (i.e., in $\hat{\vek{y}}$
direction) is a good quantum number of the DBdG Hamiltonian, and a
general eigenspinor is of the form
\begin{equation}
\Psi_{n q}(\vek{r}) = \ee^{ \frac{i\phi}{2}\, 
\sigma_0\otimes\tau_z} \,\, \ee^{i q y} \, \Phi_{nq}(x)
\: .
\end{equation}
Here $\tau_z$ is a Pauli matrix acting in Nambu space, $\sigma_0$ the
identity in spin space, and $n$ enumerates the energy (Landau) levels
for a fixed $q$. The spinors $\Phi_{nq}(x)$ are solutions of the
one-dimensional (1D) DBdG equation ${\mathcal H}(q) \, \Phi_{n q}(x) =
\varepsilon\, \Phi_{nq}(x)$, with
\begin{widetext}
\begin{equation}
{\mathcal H}(q) = \hbar v_{\text{F}} \left\{ \sigma_x\otimes 
\tau_z (-i) \partial_x + \sigma_y
\otimes \left[ \tau_z \, q + \tau_0\, \frac{e B}{\hbar}\, x \,
  \Theta(x) \right]\right\} 
-\mu \, \sigma_0\otimes\tau_z + \Delta_0\, \Theta(-x) \,\sigma_0\otimes\tau_x
\end{equation}
\end{widetext}
where the $\tau_j$ are Pauli matrices acting in Nambu space, and
$\tau_0$ the identity in Nambu space. A calculation similar to the one
performed in Ref.~\onlinecite{akh07} yields the spectrum of
Landau-level (LL) eigenenergies $\varepsilon_{n q}$ and explicit
expressions for $\Phi_{nq}(x)$ in the N and S regions (see the
Supplemental Material~\cite{suppl_material}, part A, for more details).
When $\mu\ne 0$, a finite dispersion near $q=0$ is exhibited (see
Fig.~\ref{fig:envsq} that also shows the dispersionless Landau levels
corresponding to $\mu=0$), which indicates the emergence of chiral Andreev
edge channels~\cite{hop00,akh07} that are localized near the interface
between the S and N regions.

\textit{Emergence of the Majorana mode\/}. -- 
We now examine in greater detail the states associated with the
linearly dispersing part of the $n=0$ level. A general symmetry
property~\cite{deg89} of the DBdG equation mandates that, for any
eigenstate $\Psi_{nq}$ with excitation energy $\varepsilon_{nq}$, its
particle-hole conjugate $\Xi\,\Psi_{nq}$ in Nambu space~\footnote{The
  operator $\Xi = \sigma_y \otimes \tau_y \, {\mathcal K}$ represents
  particle-hole conjugation in Nambu space~\cite{fu08}. $\tau_y$ is a
  Pauli matrix acting in Nambu space and $\mathcal K$ symbolizes
  complex conjugation.}  is also an eigenstate and has excitation
energy $-\varepsilon_{nq}$. This symmetry implies that the state with
quantum numbers $n=0$ and $q=0$ is its own particle-hole conjugate,
thus exhibiting the defining property of a Majorana
fermion~\cite{wil09,ali12,maj37}. Explicit inspection indeed verifies
the relation $\Xi\,\Psi_{00}(\vek{r}) = -i\, \Psi_{00}(\vek{r})$ (see
the Supplemental Material~\cite{suppl_material}, part B, for more
details).

While the Majorana state $\Psi_{00}(\vek{r})$ has a localized spatial
profile in the direction perpendicular to the S--N junction (i.e., the
x-direction), it is completely delocalized in the direction parallel
to the S--N junction.  To explore this Majorana mode further, we have
developed a description of single-particle excitations with $q\ne 0$
based on an adaptation of the envelope-function (EF) theory commonly
applied in semiconductor physics~\cite{*[{See, e.g., Sec.~5.6 in }]
  [{.}] ros09}. A general eigenstate of the Hamiltonian ${\mathcal
  H}(q)$ can be expressed in terms of the eigenstates at $q=0$ as
\begin{equation}
\Phi_{n q}(x) = 
\sum_{n^\prime} a_{n q}^{(n^\prime)} \, \Phi_{n^\prime 0}(x) \: .
\end{equation}
Inserting this expansion into the 1D DBdG equation and projecting onto
specific basis states $\Phi_{n^{\prime\prime}0}$ yields a set of
linear equations $\sum_{n^\prime} {\mathcal H}_{n^{\prime\prime}
  n^\prime}^{(\text{EF})}(q)\, a_{n q}^{(n^\prime)} =\varepsilon \,
a_{n q}^{(n^{\prime\prime})}$ for the coefficients, in the process
defining the general EF matrix Hamiltonian
\begin{eqnarray}\label{eq:EFAmatrix}
&& {\mathcal H}_{n^{\prime\prime} n^\prime}^{(\text{EF})}(q) = 
\varepsilon_{n^\prime 0}
\delta_{n^\prime n^{\prime\prime}} \nonumber \\ && \hspace{1cm} + 
\int_{-\infty}^{\infty} dx \,\, \left[
\Phi_{n^{\prime\prime}0}(x) \right]^\dagger \left\{ {\mathcal H}(q) - 
{\mathcal H}(0)\right\}
\Phi_{n^{\prime}0}(x) \: .
\end{eqnarray}
Specializing to the 1D DBdG Hamiltonian, we find
\begin{equation}\label{eq:qMatrix}
{\mathcal H}(q) - {\mathcal H}(0) = 
\hbar v_{\text{F}} \, q\, \sigma_y\otimes \tau_z
\equiv q\, \left[\frac{\partial {\mathcal H}(q)}{\partial q} \right]_{q=0} \: ,
\end{equation}
which is independent of spatially varying quantities (the vector and
pair potentials).  Application of the Hellmann-Feynman
theorem~\footnote{More information on the Hellmann-Feynman theorem can
be found in quantum-physics textbooks, e.g., Complement
G$_{\mathrm{XI}}$~5.b of Ref.~\onlinecite{coh77a}.} 
facilitates further calculation of the
EF matrix Hamiltonian (see the Supplemental
Material~\cite{suppl_material}, part E, for more details). With
mode-dependent velocities
\begin{equation}\label{eq:speed}
v_{n} \equiv \left. \frac{1}{\hbar}\, \frac{\partial \varepsilon_{nq}}
{\partial q} \right|_{q=0} =
v_{\text{F}} \int_{-\infty}^{\infty} dx \,\,
\left[\Phi_{n0}(x)\right]^\dagger 
\sigma_y\otimes\tau_z\, \Phi_{n0}
(x) \,\, ,
\end{equation}
we finally obtain an expression where intra-LL and inter-LL couplings
are neatly separated:
\begin{subequations}\label{eq:EFTham}
\begin{eqnarray}
H_{n^{\prime\prime} n^\prime}^{(\text{EF})}(q) &=& 
\left( \varepsilon_{n^\prime 0} + \hbar
v_{n^\prime} q \right) \delta_{n^\prime n^{\prime\prime}} + 
q\left( \varepsilon_{n^\prime 0} -
\varepsilon_{n^{\prime\prime}0} \right) 
A_{n^{\prime\prime} n^\prime} , \nonumber \\ \\
A_{n^{\prime\prime} n^\prime} 
&=& \int_{-\infty}^{\infty} dx \,\, \left[ \Phi_{n^{\prime\prime}0}(x)
\right]^\dagger \left[\frac{\partial}
{\partial q}\, \Phi_{n^{\prime}q}(x)\right]_{q=0}  \: .
\label{eq:matElem}
\end{eqnarray}
\end{subequations}

The EF approach is particularly useful to consider states with small
$q\lesssim l_B^{-1}$, as it provides a natural platform for a
perturbative treatment. To lowest (i.e., linear) order in $q$, the
dispersion of the $n=0$ Dirac-Andreev edge state is given by
$\varepsilon_{0q} = \hbar v_0 q +{\mathcal O}(q^2)$, and the
second-quantized particle annihilation operator for the state with
$n=0$ and finite small $q$ is found to be
\begin{equation}\label{eq:finQ}
\gamma_q = c_0 + q\, \sum_{n>0} A_{n0} 
\left( c_n - c_n^\dagger \right) +{\mathcal O}(q^2)\: .
\end{equation}
The $c_j$ in Eq.~(\ref{eq:finQ}) are single-particle operators
associated with the EF basis states $\Phi_{j0}(x)$, satisfying the
usual fermionic anticommutation relations, and $c_0^\dagger =
c_0$. The structure of the expression (\ref{eq:finQ}) embodies the
relation
\begin{equation}\label{eq:MajQ}
\gamma_q^\dagger = \gamma_{-q} \: ,
\end{equation}
which fundamentally arises as a consequence of particle-hole symmetry
in the DBdG equation. The identity (\ref{eq:MajQ}) is a hallmark of a
Majorana excitation~\cite{fu09}, because it enables construction of a
real-space particle operator~\footnote{The introduction of a cut-off
  scale $\Lambda\sim l_B^{-1}$ and normalization factor ${\mathcal
    N}_\Lambda$ is necessitated by the limited range in $q$ over which
  the $n=0$-mode dispersion is linear.}
\begin{equation}\label{eq:realMaj}
\gamma_\Lambda(y) = 
{\mathcal N}_\Lambda \int_{-\Lambda}^\Lambda dq \,\, \ee^{i q y}\,\gamma_q
\end{equation}
that satisfies the condition 
$\gamma_\Lambda(y)^\dagger = \gamma_\Lambda(y)$.

Majorana fermions are often also called `half-fermions' because an
ordinary (complex, Dirac) electron can be considered to be made up of
two Majorana (real) fermion degrees of freedom with perfectly
synchronized dynamics. Thus the emergence of independent Majorana
excitations in an electronic system is often due to the
fractionalization of an ordinary electron state, brought about by
complex correlations in a material's electronic ground
state. Naturally, a fractionalization scenario always generates pairs
of free Majorana excitations, and the physical separation of the
constituents of the pair has to be ensured to avoid their
`recombination' into an ordinary electron.  The breaking of
time-reversal symmetry due to the applied magnetic field eliminates
the possibility for there to be a doubly occupied zero-energy state at
$x=0$ with guiding center $q=0$. As particle-hole symmetry is
preserved, such a state has to be a chiral Majorana excitation. There
is no partner-complement along the NS-interface $x=0$ to form a Dirac
fermion, hence this excitation will be stable (e.g. against disorder)
as long as a zero-energy state can exist.

\begin{figure}[b]
\includegraphics[angle=0,scale=1.2]{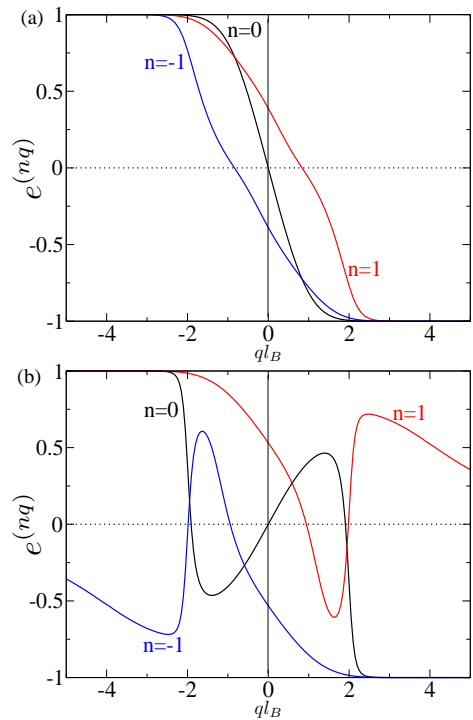}
\caption{\label{fig:effCharge} (Color online)
Effective electric charge $e^{(nq)}$ of the single-particle
excitations for the parameters used in Figs.~\ref{fig:envsq}a and b,
numerically calculated using the normal-side wavefunctions in
Eq.~(\ref{eq:effCharge}). 
For both sets of parameters, $e^{(00)}=0$ as expected for a
Majorana state. Notice also the linear variation of $e^{(nq)}$ for small
$q\lesssim l_B^{-1}$.}
\end{figure}

\textit{Electric charge of Dirac-Andreev edge states\/}. As is the
case for Bogoliubov quasiparticles in a conventional
superconductor~\cite{deg89,kiv90,nay01,fuj08,par12}, the Dirac-Andreev
edge states are coherent superpositions of particles and holes in
Nambu space. Consequently, their effective electric charge generally
has a non-quantized value
\begin{equation}\label{eq:effCharge}
e^{(\Psi)} = \int d^2 r \, \left[\Psi(\vek{r})\right]^\dagger 
\sigma_0\otimes \tau_z \, \Psi(\vek{r})
\end{equation}
that depends on the particular state. Figure~\ref{fig:effCharge} shows
the numerically calculated charges $e^{(nq)}$ of states from the bands
$n=0,\pm 1$ given in Fig.~\ref{fig:envsq}.

\begin{figure}
\includegraphics[angle=0,scale=0.75]{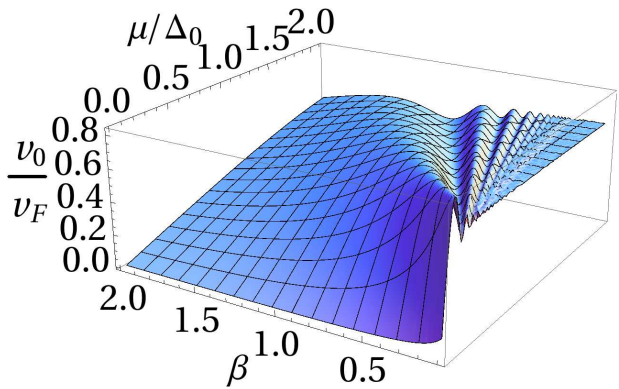}
\caption{\label{fig:vmvsbm} (Color online)
Dependence of the Majorana-mode velocity $v_0$ on system parameters.
$\beta=\sqrt{\hbar |e B|} v_F/\Delta_0$ measures the magnetic-field
strength, $\mu$ is the superconductor's chemical potential, and
$\Delta_0$ the magnitude of the superconducting pair
potential. 
}
\end{figure}

Using the EF theory discussed above, we can express the charge of
modes with small $q\ne0$ as
\begin{subequations}
\begin{equation}
e^{(nq)} = e_{nn} + 2 q\,\, 
\Re{\mathrm e}\sum_{n'\ne n} A_{n' n} \, e_{n' n} + {\mathcal O}(q^2) \: ,
\end{equation}
with the basis-function-related charge parameters
\begin{equation}
e_{n' n} = \int dx \, \left[ \Phi_{n'0}(x) \right]^\dagger 
\sigma_0\otimes \tau_z \, \Phi_{n0}(x)
\: .
\end{equation}
\end{subequations}
By definition, $e_{n n}\equiv e^{(n0)}$ is the effective charge of the
EF basis state with quantum number $n$. The $q$-proportional coupling
of basis states results in a linear $q$ dependence of $e_{nq}$ for
$q\lesssim l_B^{-1}$, in agreement with Fig.~\ref{fig:effCharge}.

For the $n=0$ (Majorana) mode, $e^{(0,-q)}=-e^{(0q)}$ and, as
expected, the state at $q=0$ has zero effective charge. A general
superposition $\Psi_0(\vek{r})\equiv \sum_{|q| <\Lambda} \alpha_q \,
\Psi_{0q}(\vek{r})$ of Majorana-mode states has total electric charge
\begin{equation}\label{eq:majCharge}
e^{(\Psi_0)} = \frac{\av{\varepsilon}}{\hbar v_0}\, 
\left.\frac{\partial e^{(0q)}}{\partial q}\right|_{q=0}
\end{equation}
proportional to its mean energy $\av{\varepsilon} \equiv \sum_{|q| <\Lambda} |
\alpha_q|^2\,\varepsilon_{0q}$.  Thus the Majorana mode's energy relaxation
will be associated with charge conversion, which is a general feature
for bogolon excitations in superconductors~\cite{kiv90,par12}.
Furthermore, as the superposition associated with the
real-space Majorana-particle operator $\gamma_\Lambda(y)$ involves
states with $\pm q$ in equal measure, its total charge vanishes as is
expected for a true Majorana fermion. 
An experimental verification of this property can possibly be done along
similar lines as used in recent experiments on edge-magnetoplasmonic
excitations in graphene \cite{petkovic2012} and open up possibilities
for electrically manipulating this mode.

\textit{Tailoring Majorana-mode properties\/}. -- An interesting
feature of this realization of a Majorana mode is that its velocity
can be tuned via the external magnetic field. This enables a
potentially easier route towards tailoring this characteristic property
than changing the magnetization of a ferromagnetic insulator
in structures of the type discussed in Ref.~\onlinecite{fu09}. The
mode-velocity dependence on the external magnetic field and the chemical
potential is shown in Fig.~\ref{fig:vmvsbm}.  Furthermore, the charge within
the $n=0$ Landau level can be controlled via the chemical potential. This is
evident by comparing the slopes of $e^{(nq)}$ at $q=0$ in the two
panels in Fig.~\ref{fig:effCharge}.

\textit{Conclusions\/}. -- 
We have demonstrated that a Majorana mode can be realized using the
interplay of Landau quantization and Andreev reflection in a
superconductor--topological-insulator hybrid structure. This mode is
robust against disorder due to its chiral nature and associated
uniqueness of the zero-energy state that is localized at the
interface, and its velocity is tunable by adjusting the strength of
the magnetic field.  We have developed a theory to describe the
coupling of this mode to other Dirac-Andreev edge states and showed
how this coupling results in an energy-dependent electric charge for
Majorana-mode states. This feature should enable creation and
manipulation of such excitations using external gate voltages.

\textit{Acknowledgments\/}. -- 
We acknowledge financial support by the Swiss SNF, 
the NCCR Nanoscience, and the NCCR Quantum Science and Technology.

\appendix

\section{\label{app:DBdG} DBdG eigenspinors in the N region}

To simplify the notation we measure all energies in units of $\hbar
v_F/l_B$ and lengths in units of $l_B$.  The superconducting pair
potential is in general, a complex quantity $\Delta=\Delta_0
e^{i\phi}$.  To have a transparent picture of the relative
phase between the electronic and the hole excitations we make the ansatz
$\Psi (x, y)=e^{iqy} ( u_{\uparrow}(x) e^{i\phi/2}, u_{\downarrow}(x) 
e^{i\phi/2}, v_{\downarrow}(x) e^{-i\phi/2}, -v_{\uparrow}(x) e^{-i\phi/2})^T$
and obtain a one-dimensional DBdG equation whose solutions that decay
for $x\rightarrow \infty$ in region $x>0$ have the form
\begin{widetext}
\begin{equation}
\Psi(x,y)=e^{iqy}\left(\begin{array}{c}
                             -i C_e e^{-\frac{(x + q)^2}{2}}(\mu +\varepsilon)H_{(\mu +\varepsilon)^2/2 -1}(x +q) e^{i\phi/2} \\
                             C_e e^{-\frac{(x + q)^2}{2}} H_{(\mu +\varepsilon)^2/2}(x +q) e^{i\phi/2} \\
                             C_h e^{-\frac{(x - q)^2}{2}} H_{(\mu -\varepsilon)^2/2}(x -q) e^{-i\phi/2} \\
                             -i C_h e^{-\frac{(x - q)^2}{2}}(\mu -\varepsilon)H_{(\mu -\varepsilon)^2/2 -1}(x -q) e^{-i\phi/2}
                            \end{array}
\right),
\label{eq:wfn}
\end{equation}
\end{widetext}
 with $H_{\alpha}(x)$ denoting the Hermite function~\cite{akh07}. The
complete solution is then obtained by demanding the conservation of
particle current along the boundary. Similar to Ref.~\cite{akh07} we
find 
\begin{equation}
C_e=\frac{-iC_h(\mu-\varepsilon)H_{(\mu -\varepsilon)^2/2 -1}(-q)}{H_{(\mu +\varepsilon)^2/2}(q) \frac{\varepsilon}{\Delta_0}+(\mu +\varepsilon)H_{(\mu +\varepsilon)^2/2 -1}(q) \sqrt{1-\frac{\varepsilon^2}{\Delta_0^2}}}, 
\label{eq:cech}
\end{equation}
and the dispersion relation is given by the solutions of
\begin{eqnarray}
f_{\mu+\varepsilon}(q)-f_{\mu-\varepsilon}(-q)&=&\frac{\varepsilon(f_{\mu+\varepsilon}(q)f_{\mu-\varepsilon}(-q)+1)}{\sqrt{\Delta_0^2-\varepsilon^2}}, \nonumber \\
f_\alpha(q)&=&\frac{H_{\alpha^2/2}(q)}{\alpha  H_{\alpha^2/2 -1}(q)}.
\label{eq:eigenvalue}
\end{eqnarray}

The solutions $\varepsilon_n(q)$ of Eq.~(\ref{eq:eigenvalue}) can be
labeled with a Landau level (LL) index $n=0, \pm 1, \pm 2,
\hdots$. Figure 2(a) in the main text shows the zeroth LL
($n=0$), along with the neighboring $n=\pm1$ LLs when
$\mu=0.4 \hbar v_F/l_B$ and $\Delta=5 \hbar v_F/l_B$. The dotted lines
at 0, $\pm \sqrt{2}$ show the dispersionless LLs
corresponding to $\mu=0$. The various levels saturate at
$\sqrt{2}(\hbar v_F/l_B)$sgn$(n)\sqrt{\mid n \mid} - \mu$ for $q
\rightarrow-\infty$ and $\sqrt{2}(\hbar v_F/l_B)$sgn$(n)\sqrt{\mid n
  \mid} + \mu$ for $q \rightarrow\infty$.  This suggests that an
interesting regime can be reached by increasing $\mu$, so that
$n=\pm1$ levels start contributing to the low energy excitations of
the system. When the chemical potential $\mu$ is finite (as measured 
from the charge-neutrality point of the free Dirac system), the LLs acquire a dispersion 
around $q=0$ that signals the existence of Andreev edge excitations~\cite{hop00,akh07}.

\section{\label{sec:majCheck} Verification of $\Psi_{00}$ as a Majorana state}

For $\varepsilon=0$ and $q=0$ we get $C_h=iC_e$, and this zero energy
state can be expressed as
\begin{equation}
\Psi_{00}(x)=C_e e^{-\frac{x^2}{2}}\left(\begin{array}{c}
            -i \mu H_{\mu^2/2 -1}(x)e^{i\phi/2} \\
              H_{\mu^2/2}(x)e^{i\phi/2} \\
            i  H_{\mu^2/2}(x)e^{-i\phi/2} \\
              \mu H_{\mu^2/2 -1}(x)e^{-i\phi/2}
           \end{array}\right).
\label{eq:majn_b}
\end{equation}

The particle-hole conjugation operator is given by $\Xi=\sigma_y\tau_y
\mathcal{K}$ ~\cite{fu08}, where $\sigma$ and $\tau$ are Pauli
matrices acting on spin and particle-hole space respectively, and
$\mathcal{K}$ is complex conjugation. Straightforward verification
implies $\Xi\Psi_{00} = -i\Psi_{00}$, and
$\Xi\Xi\Psi_{00}=\Psi_{00}$. Hence the state $\Psi_{00}$ is a Majorana
fermion.

\section{DBdG eigenspinors in the S region}

In the superconducting region $x<0$ the solutions of
Eq.~(1) in the main text take the form~\cite{beenakker2006}
\begin{widetext}
\begin{equation}
\Psi(x,y) =A e^{iqy} e^{i k_0 x} e^{\kappa x}\left(\begin{array}{c}
                             e^{-i\theta} e^{i\phi/2}\\
                             e^{i\gamma}e^{-i\theta} e^{i\phi/2}\\
                             e^{-i\phi/2}\\
                             e^{i\gamma} e^{-i\phi/2} 
                             \end{array}
\right) + B e^{iqy} e^{-i k_0 x} e^{\kappa x}\left(\begin{array}{c}
                             e^{i\theta} e^{i\phi/2}\\
                             -e^{-i\gamma}e^{i\theta} e^{i\phi/2}\\
                             e^{-i\phi/2} \\
                             -e^{-i\gamma} e^{-i\phi/2} 
                             \end{array}
\right),  
\label{eq:wfs}
\end{equation}
\end{widetext}
where the parameters $\theta, \gamma, k_0, \kappa$ are defined by
\begin{equation}
\theta=\begin{cases}\cos^{-1}\frac{\varepsilon}{\Delta_0} & \text{if }  \varepsilon < \Delta_0 \\
-i\cosh^{-1}\frac{\varepsilon}{\Delta_0} & \text{if }  \varepsilon > \Delta_0
\end{cases},
\label{eq:wfsparam}
\end{equation}
and $\gamma=\sin^{-1}\frac{q}{\mu}, k_0=\sqrt{\mu^2 - q^2}, \kappa=\frac{\mu\Delta_0}{k_0}\sin(\theta).$

It should be noted that the solutions of the DBdG Hamiltonian given by
Eq.~(\ref{eq:wfs}), are approximate solutions valid in the regime
$\mu\gg \Delta_0,\varepsilon$ for all values of $q$.  We are
interested in the eigenstate at $q=0$ where Eq.~(\ref{eq:wfs})
represents an exact solution.

The coefficients $A, B, C_e$, and $C_h$ can be obtained by demanding
the continuity of the wavefunction at $x=0$~\cite{akh07}. Due to
translational invariance in the $y$ direction, the wavefunction
matching can be done for each $q$ separately. Thus we obtain (for
$q=0$ and $\varepsilon=0$)
\begin{eqnarray}
-i A  + i B  &=&  C_e (-i)\mu H_{\mu^2/2 -1}(0), \nonumber \\
-i A - i B &=& C_e H_{\mu^2/2}(0), \nonumber \\
A + B &=& C_h H_{\mu^2/2}(0), \nonumber \\
A - B &=& C_h (-i)\mu H_{\mu^2/2 -1}(0).
\label{eq:wfcoef}
\end{eqnarray}
Once again we obtain $C_h=iC_e$. The eigenspinor for the Majorana
state for $x>0$ is given by Eq.~(\ref{eq:majn_b}), while for $x<0$ it
is
\begin{equation}
\Psi_{00}(x)= e^{\Delta_0 x}\left(\begin{array}{c}
            (-i A e^{i\mu x} + i B e^{-i\mu x})e^{i\phi/2} \\
            (-i A e^{i\mu x} - i B e^{-i\mu x})e^{i\phi/2} \\
            (A e^{i\mu x} + B e^{-i\mu x})e^{-i\phi/2} \\
            (A e^{i\mu x} - B e^{-i\mu x})e^{-i\phi/2}
           \end{array}\right),
\label{eq:majs}
\end{equation}
where 
\begin{eqnarray}
A &=& \frac{C_e}{2} (\mu H_{\mu^2/2 -1}(0)+i H_{\mu^2/2}(0)), \nonumber \\
B &=& \frac{C_e}{2} (-\mu H_{\mu^2/2 -1}(0)+i H_{\mu^2/2}(0)).
\label{eq:ab}
\end{eqnarray}

Once again we can verify that the state is its own particle-hole
conjugate. Thus we have obtained the state representing the Majorana
fermion in first-quantized form in the entire space.

\section{\label{sec:Heff_anal}Calculation of the envelope-function 
Hamiltonian using Eq.~(5)}

Using the explicit expressions for the eigenstates given by
Eqs.~(\ref{eq:wfn}) and (\ref{eq:wfs}) we can write the EF matrix
Hamiltonian as
\begin{equation}
{\mathcal H}_{n^{\prime\prime} n^\prime}^{(\text{EF})}(q) 
= \varepsilon_{n^\prime 0}\delta_{n^\prime n^{\prime\prime}}
+\hbar v_Fq\mathcal{B}_{n^{\prime\prime} n^\prime},
\end{equation}
where 
\begin{widetext}
\begin{eqnarray}
{\mathcal B}_{n^{\prime\prime} n^\prime}&=&\sqrt{N(\varepsilon_{n^\prime 0})N(\varepsilon_{n^{\prime\prime} 0})} [ 
2^{\frac{(\mu + \varepsilon_{n^\prime 0})^2+(\mu + \varepsilon_{n^{\prime\prime} 0})^2}{2}-1}\pi\left(g(\varepsilon_{n^\prime 0},\varepsilon_{n^{\prime\prime} 0})
+g(\varepsilon_{n^{\prime\prime} 0},\varepsilon_{n^\prime 0})\right) \nonumber \\
&+& G(\varepsilon_{n^\prime 0},\varepsilon_{n^{\prime\prime} 0})
2^{\frac{(\mu + \varepsilon_{n^\prime 0})^2+(\mu + \varepsilon_{n^{\prime\prime} 0})^2}{2}-1}\pi\left(g(-\varepsilon_{n^\prime 0},-\varepsilon_{n^{\prime\prime} 0})
+g(-\varepsilon_{n^{\prime\prime} 0},-\varepsilon_{n^\prime 0})\right) ].
\end{eqnarray}
\end{widetext}
The various functions $N(\eta), g(\eta,\zeta)$, and
$G(\eta,\zeta)$ can be expressed using Gamma functions:
\begin{widetext}
 \begin{eqnarray}
  N(\eta)&=&\frac{2^{\frac{(\mu+\eta)^2}{2}}\sqrt{\pi}}{4}\left(\frac{\psi(\frac{1}{2}-\frac{(\mu+\eta)^2}{4})-\psi(-\frac{(\mu+\eta)^2}{4})}{\Gamma(-\frac{(\mu+\eta)^2}{2})}+
\frac{(\mu+\eta)^2}{2}\frac{\psi(1-\frac{(\mu+\eta)^2}{4})-\psi(\frac{1}{2}-\frac{(\mu+\eta)^2}{4})}{\Gamma(1-\frac{(\mu+\eta)^2}{2})}\right) \nonumber \\
&+& \frac{2^{\frac{(\mu-\eta)^2}{2}}\sqrt{\pi}}{4}\left(\frac{\psi(\frac{1}{2}-\frac{(\mu-\eta)^2}{4})-\psi(-\frac{(\mu-\eta)^2}{4})}{\Gamma(-\frac{(\mu-\eta)^2}{2})}+
\frac{(\mu-\eta)^2}{2}\frac{\psi(1-\frac{(\mu-\eta)^2}{4})-\psi(\frac{1}{2}-\frac{(\mu-\eta)^2}{4})}{\Gamma(1-\frac{(\mu-\eta)^2}{2})}\right)\nonumber \\
&+& \frac{\left((\mu-\eta)^2\Gamma^2(\frac{1}{2}-\frac{(\mu-\eta)^2}{4})+4\Gamma^2(1-\frac{(\mu-\eta)^2}{4})\right)\left(
2\eta\Gamma(1-\frac{(\mu+\eta)^2}{4})+\Delta_0\sqrt{1-\frac{\eta^2}{\Delta_0^2}}(\eta+\mu)\Gamma(\frac{1}{2}-\frac{(\mu+\eta)^2}{4})
\right)^2}
{4\Delta_0^3\sqrt{1-\frac{\eta^2}{\Delta_0^2}}(\mu-\eta)^2\Gamma^2(\frac{1}{2}-\frac{(\mu-\eta)^2}{4})\Gamma^2(1-\frac{(\mu+\eta)^2}{4})}
\nonumber \\
g(\eta,\zeta)&=& \frac{\mu+\zeta}{\frac{(\mu+\eta)^2}{2}-\frac{(\mu+\zeta)^2}{2}+1}\left(\frac{1}{\Gamma(\frac{1}{2}-\frac{(\mu+\eta)^2}{4})\Gamma(\frac{1}{2}-\frac{(\mu+\zeta)^2}{4})}-
\frac{1}{\Gamma(1-\frac{(\mu+\zeta)^2}{4})\Gamma(-\frac{(\mu+\eta)^2}{4})}
\right) \nonumber \\
G(\eta,\zeta)&=&\left(\frac{H_{\frac{(\mu+\eta)^2}{2}}(0)\frac{\eta}{\Delta_0}+(\mu+\eta)H_{\frac{(\mu+\eta)^2}{2}-1}(0)\sqrt{1-\frac{\eta^2}{\Delta_0^2}}}{(\mu-\eta)H_{\frac{(\mu-\eta)^2}{2}-1}(0)}\right)
\left(\frac{H_{\frac{(\mu+\zeta)^2}{2}}(0)\frac{\zeta}{\Delta_0}+(\mu+\zeta)H_{\frac{(\mu+\zeta)^2}{2}-1}(0)\sqrt{1-\frac{\zeta^2}{\Delta_0^2}}}{(\mu-\zeta)H_{\frac{(\mu-\zeta)^2}{2}-1}(0)}\right)\nonumber, 
 \end{eqnarray}
\end{widetext}
where $\psi$ and $\Gamma$ denote the Digamma function and the Gamma
function respectively~\cite{gra94}.

 \section{\label{sec:HFtheorem}Calculation of the envelope-function 
Hamiltonian using the  Hellmann-Feynman theorem}

The r.h.s\ of Eq.~(6) in the main text lends itself to the application
of the Hellman-Feynman theorem. Using a straightforward Dirac
notation, we have
\begin{widetext}
\begin{subequations}
\begin{eqnarray}
\bra{n^{\prime\prime}0} \left[\frac{\partial {\mathcal H}(q)}{\partial q} \right]_{q=0}
\ket{n^\prime 0} &=& \bra{n^{\prime\prime}0} \left[ \frac{\partial}{\partial q} \left\{
{\mathcal H}(q) \ket{n^\prime q}\right\} - {\mathcal H}(q) \, \frac{\partial}{\partial q}
\ket{n^\prime q}\right]_{q=0} \: , \\
&=& \bra{n^{\prime\prime}0} \left[\frac{\partial \varepsilon_{n^\prime q}}{\partial q}
\right]_{q=0} \ket{n^\prime 0} + \bra{n^{\prime\prime}0}  \left\{ \varepsilon_{n^\prime 0}
- {\mathcal H}(0)\right\} \left[ \frac{\partial}{\partial q} \ket{n^\prime q}\right]_{q=0} \: , \\
&=& \left[\frac{\partial \varepsilon_{n^\prime q}}{\partial q} \right]_{q=0} \,\delta_{n^{\prime
\prime} n^\prime} + \left( \varepsilon_{n^\prime 0} - \varepsilon_{n^{\prime\prime}0} \right)
\bra{n^{\prime\prime}0} \left[ \frac{\partial}{\partial q} \ket{n^\prime q}\right]_{q=0} \: .
\label{eq:HFresult}
\end{eqnarray}
\end{subequations}
\end{widetext}
The expression given in Eq.~(\ref{eq:HFresult}) neatly separates the
intra-band and inter-band matrix elements in the first and second
terms, respectively. Inserting into Eq.~(5) in the main text and using
the definition Eq.~(7) in the main text yields Eqs.~(8) in the main text.

Naturally, the full matrix $H_{n^{\prime\prime} n^\prime}(q)$ is not
tractable. However, we are usually only interested in the bands in the
vicinity of the chemical potential, as the relevant dynamics of the
material will take place mostly in these. The coupling to more remote
bands will affect the parameters in any reduced-band model (e.g., the
effective mass in conventional semiconductor band models, or the
velocity of the edge modes in our current system of interest). The
systematic way to derive a reduced-band model is L\"owdin
partitioning (see, e.g., Appendix~B in Ref.~\cite{win03} for a
pedagogical introduction).

\begin{figure}
\includegraphics[angle=0,scale=0.6,clip]{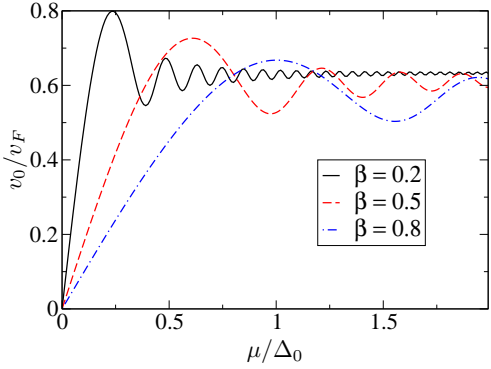}
\caption{Velocity of the Majorana mode along $\hat{y}$ direction as a
  function $\mu/\Delta_0$ for different values of the magnetic field. $\beta\equiv\sqrt{\hbar|eB|}v_F/\Delta_0$ defines a dimensionless measure of the magnetic field. 
See the text for further details.
}
\label{fig:vm2dm}
\end{figure}

\begin{figure}
\includegraphics[angle=0,scale=0.6,clip]{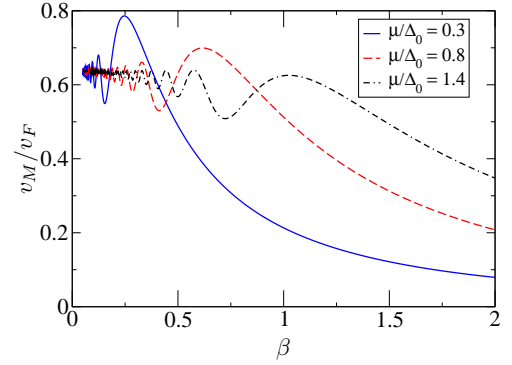}
\caption{Velocity of the Majorana mode along $\hat{y}$ direction as a
  function of $\beta$ for different values of $\mu/\Delta_0$.  See the text for further details.}
\label{fig:vm2db}
\end{figure}

\section{Majorana mode coupled to nearest bands as an illustration of 
envelope-function theory}

We will now consider the coupling of the Majorana mode ($n=0$) to the
nearest LLs (with $n=\pm 1$). The EF matrix Hamiltonian for this
situation is given by (we use the notation $\bar 1 \equiv -1$)
\begin{equation}
\tilde{\mathcal H} = \left( \begin{array}{ccc} \hbar v_0 q & \varepsilon_1 A_{01} q & 
-\varepsilon_1 A_{0\bar 1} q \\ -\varepsilon_1 A_{10} q & \varepsilon_1 + \hbar v_1 q &
-2\varepsilon_1 A_{1\bar 1} q \\ \varepsilon_1 A_{\bar 10} q & 2\varepsilon_1 A_{\bar 11} q
& -\varepsilon_1 + \hbar v_1 q  \end{array}\right) \: ,
\end{equation}
where $\varepsilon_1\equiv \varepsilon_{10} = -\varepsilon_{\bar
  10}$. Straightforward diagonalization yields eigenvalues
$\varepsilon_{\pm 1 q} = \pm \varepsilon_1 + \hbar v_1 q+{\mathcal
  O}(q^2)$, $\varepsilon_{0q} = \hbar v_0 q+{\mathcal O}(q^2)$ and
associated eigenstates $\chi_{nq} \equiv
(a_{nq}^{(0)},a_{nq}^{(1)},a_{nq}^{(\bar 1)})^T$
\begin{equation}
\chi_{0q} = \left(\begin{array}{c} 1 \\ A_{10} q \\ A_{\bar 10} q \end{array}\right) ,
\chi_{1q} = \left(\begin{array}{c} A_{01}q \\ 1 \\ A_{\bar 11} q \end{array}\right) ,
\chi_{\bar 1q} = \left(\begin{array}{c} A_{0\bar 1} q \\ A_{1\bar 1} q \\ 1 \end{array}\right) ,
\end{equation}
also with accuracy of ${\mathcal O}(q^2)$. Thus, up to linear order in
$q$, the inter-band coupling does not affect the linear dispersion of
each mode, but an admixture between the different modes is present in
the eigenstates.

We now consider the second-quantized representation of the
Dirac-Andreev edge states.  Based on the above discussion, we can
write the annihilation operator for the state with wave number $q$ in
the $n=0$ band as
\begin{subequations}
\begin{equation}
\gamma_q = c_0 + q\left( A_{10}\, c_1 + A_{\bar 10}\, c_{\bar 1}\right) \: .
\end{equation}
Here $c_n$ are the second-quantized operators associated with the
$q=0$ states in band $n$, i.e., our basis states. By the symmetry
inherent to the Dirac-BdG equations, we have $c_0 = c_0^\dagger$ and
$A_{\bar 10}\, c_{\bar 1}\equiv - A_{10}\, c_1^\dagger$ and thus find
\begin{equation}
\gamma_q = c_0 + q \, A_{10} \big( c_1 - c_1^\dagger \big) \: .
\end{equation}
\end{subequations}

\section{Tuning the system properties via the external magnetic field}

It should be noted that all the energy scales and momentum scales in
Eq.~(\ref{eq:eigenvalue}) are defined using the magnetic length thus
removing any explicit dependence upon the external magnetic field.  If
we are interested in tuning the properties of our system by changing
the external magnetic field, it is useful to define another set of
dimensionless units where we can measure the strength of the magnetic
field, this is achieved by measuring all the energies in units of the
superconducting gap.  We define
$\tilde{\mu}\equiv\frac{\mu}{\Delta_0\beta}$,
$\tilde{\varepsilon}\equiv\frac{\varepsilon}{\Delta_0\beta}$,
$\tilde{q}\equiv\frac{\hbar v_Fq}{\Delta_0\beta}$ where $\beta$ 
is a dimensionless measure of the magnetic field 
($\beta\equiv\frac{\sqrt{\hbar |e B|} v_F}{\Delta_0}$). 
Using this Eq.~(\ref{eq:eigenvalue}) can be rewritten as
\begin{equation}
f_{\tilde{\mu}+\tilde{\varepsilon}}(\tilde{q})
-f_{\tilde{\mu}-\tilde{\varepsilon}}(-\tilde{q})=
\frac{\beta\tilde{\varepsilon}(
f_{\tilde{\mu}+\tilde{\varepsilon}}(\tilde{q})
f_{\tilde{\mu}-\tilde{\varepsilon}}(-\tilde{q})+1)}
{\sqrt{1-\beta^2\tilde{\varepsilon}^2}}\:,
\label{eq:eigenvaluenewunits}
\end{equation}
where the function $f_\alpha(q)$ is the same as defined
earlier. Equation~(\ref{eq:eigenvaluenewunits}) can be solved for
$\tilde{\varepsilon}$.  The numerical results for the velocity of the
Majorana mode as a function of $\beta$ and $\mu/\Delta_0$ is shown in
Fig.~4 in the main text. In this supplemental material Fig.~\ref{fig:vm2dm} 
shows $v_0/v_F$ as a function of $\mu/\Delta_0$ for different values of 
$\beta$ and similarly Fig.~\ref{fig:vm2db} shows $v_0/v_F$ as a function 
of $\beta$ for different values of $\mu/\Delta_0$ just to illustrate the oscillating behavior
of $v_0/v_F$ as seen in Fig.~4 in the main text.


%

\end{document}